\documentclass{aa}
\usepackage{times}
\usepackage{graphics}
\usepackage[dvips]{epsfig}
\begin{document}
\thesaurus{04(12.07.1,12.04.1,08.12.2,10.08.1)}
\title{Where are the binary source galactic microlensing events?}
\author{M. Dominik\thanks{\emph{Present address:} Space Telescope Science Institute, 3700 San Martin Drive, Baltimore, MD 21218, USA (dominik@stsci.edu)}}
\institute{Institut f\"ur Physik, Universit\"at Dortmund, D-44221 Dortmund, Germany}
\date{Received ; accepted}
\maketitle
\begin{abstract}
Though there have been some galactic microlensing events which show a clear
signature of a binary lens, no event has yet been claimed as due to lensing
of a binary source. Here I argue that this may be due to the fact that most
of the binary source events show light curves which can be fitted with the 
simpler model of a blended single source.

\end{abstract}

\section{Introduction}
Among the galactic microlensing events detected by the 4 observing groups
EROS (Aubourg et al. \cite{aubourg}), MACHO (Alcock et al. \cite{alcock1}, 
1997a,b\nocite{alcock3}\nocite{alcock4}), 
OGLE (Udalski et al. 1994a-e)\nocite{uda1}\nocite{uda2}\nocite{uda3}
\nocite{uda4}\nocite{uda5}, and DUO (Alard et al.~1995a,b\nocite{alard2}\nocite{amg}),
some events show the characteristics of a binary
lens, namely MACHO LMC\#1 (Dominik \& Hirshfeld \cite{dohi1}, \cite{dohi2}), OGLE\#7
(Udalski et al. \cite{uda4}), DUO\#2 (Alard et al. \cite{amg}), 
MACHO LMC\#9 (Bennett et al. \cite{bennett}),
MACHO Bulge 95-12 (Pratt et al. \cite{pratt}), and MACHO Bulge 96-3 
(Stubbs et al. \cite{stubbs}).
In contrast, no event has been claimed to involve
a binary source, though Griest \& Hu (\cite{grihu}) have predicted that
around 10~\% of the events should involve features
of a binary source. Given this situation, one may clearly pose the question
where the binary source microlensing events are.

Here I argue that the lack of claimed binary source microlensing events
may be due to the fact that most of the light curves for such events
can successfully be explained with the simpler model of a blended 
single source. In fact, such a model is successful for the event
OGLE\#5 (Dominik \cite{domthesis}; Alard \cite{alard}) and most of the DUO events also
involve blending (Alard \cite{alard}). In this paper, I discuss
also a model with a binary source for OGLE\#5 and compare it
with the model involving a blended single source.

\section{Point-mass lens and point source}
\label{SSmodel}
For a lens at a distance $D_\mathrm{d}$ from the observer, a
source at a distance $D_\mathrm{s}$ from the observer, and $D_\mathrm{ds}$
the lens-source distance, the Einstein radius for a lens of
mass $M$ is given by
\begin{equation}
r_\mathrm{E} = \sqrt{\frac{4GM}{c^2} \, \frac{D_\mathrm{d} D_\mathrm{ds}}{D_\mathrm{s}}}\,.
\end{equation}

Let the lens move on a straight line with a velocity $v_{\perp}$ 
transverse to the line-of-sight 
observer-source, so that it moves 
one Einstein radius in the lens plane in the time
$t_\mathrm{E} = \frac{r_\mathrm{E}}{v_{\perp}}$.\footnote{Note that this
is equivalent to letting the source projected onto the lens plane move
with $v_{\perp}$ into the opposite direction. Further note that if the 
fixed source is on the right side of the lens trajectory, a corresponding
fixed lens is on the right side of the source trajectory.}
Let $t_\mathrm{max}$ denote the time at the closest approach and
$u_\mathrm{min} = r_\mathrm{min}/r_\mathrm{E}$ the impact parameter at
$t_\mathrm{max}$ in units of Einstein radii.
For the impact parameter at time $t$ one obtains
\begin{equation}
u = \sqrt{u_\mathrm{min}^2 + [p(t)]^2}\,,
\end{equation}
where 
\begin{equation}
p(t) = \frac{t-t_\mathrm{max}}{t_\mathrm{E}}\,.
\end{equation} 
The light amplification for a point source and a point-mass lens is given by
\begin{equation}
A_\mathrm{SS}(u(t)) = \frac{u^2 + 2}{u \sqrt{u^2+4}}\,.
\end{equation}

\section{Point-mass lens and binary source}
\label{BSmodel}
For a binary source, according to Griest \& Hu (\cite{grihu}),
two values of $t_\mathrm{max}$ and $u_\mathrm{min}$ 
defining the closest approach to the first and to the second 
source object are used, which imply two functions $u_1(t)$ and $u_2(t)$. With 
$L_1$ and $L_2$ being the luminosities of the two parts and the
luminosity offset ratio
\begin{equation}
\omega = \frac{L_2}{L_1 + L_2} \,,
\end{equation}
the light amplification is given by
\begin{equation}
A_\mathrm{BS}(u_1(t),u_2(t)) = (1-\omega) A_\mathrm{SS}(u_1) + 
	\omega A_\mathrm{SS}(u_2)\,,\label{binary}
\end{equation}
so that the light curve is a superposition of two light curves for 
point sources behind point-mass lenses.

Since the two components of the binary source may be located on the same
side or on opposite sides of the 
lens trajectory without changing the distance
functions $u_1(t)$ and $u_2(t)$, two different physical configurations
producing the same light curve exist. I call the case where the components
are on the same side the {\em cis}-configuration, the case where
the components are on opposite sides the {\em trans}-configuration.

Let $2 \rho$ denote the distance between the closest approaches of the lens
to the two components and $2 \lambda$ denote the distance between the
source components, both measured in projected Einstein radii 
$r_E' = \frac{D_\mathrm{s}}{D_\mathrm{d}}\,r_\mathrm{E}$.
It follows that
\begin{equation}
\rho = \frac{t_\mathrm{max,2}-t_\mathrm{max,1}}{2 t_\mathrm{E}}\,.
\end{equation}
The angle $\beta$ between the direction from source component 1 to source component
2 and the lens trajectory is given by
\begin{eqnarray}
\beta  & = &\arctan \frac{u_\mathrm{min,2} \pm u_\mathrm{min,1}}{2 \rho}
\nonumber \\ & =&  \arctan \left(\frac{u_\mathrm{min,2} \pm u_\mathrm{min,1}}{ 
  t_\mathrm{max,2}-t_\mathrm{max,1}} \, t_\mathrm{E}\right)
 \,,
\end{eqnarray}
where the upper sign refers to the trans-configuration and the lower sign
refers to the cis-configuration.
The half-distance between the components follows as
\begin{equation}
\lambda = \rho / \cos \beta\,.
\end{equation}

\section{Comparison of binary source and blended single source}
In contrast to the light amplification for a binary source (Eq.~(\ref{binary})),
one obtains for a blended single source
\begin{equation}
A_\mathrm{blend} = f\,A_\mathrm{SS}(u) + 1-f\,,\label{blend}
\end{equation}
where the blending parameter $f$ gives the
contribution of the light of the source at unlensed state 
to the total light 
(source and component which does not undergo any lensing).

In the case of large $u_1$, one has $A_\mathrm{SS}(u_1) \approx 1$ and 
therefore
\begin{equation}
A_\mathrm{BS}(u_1(t),u_2(t)) \approx 1-\omega + \omega A_\mathrm{SS}(u_2)\,,
\end{equation}
i.e. the light curve for a binary source approaches that for a blended single
source (object 2), and $\omega \approx f$, where the blended single source is the exact
limit of the binary source for $u_1 \to \infty$. Similarly, for large $u_2$,
one has $A_\mathrm{SS}(u_2) \approx 1$ and
\begin{equation}
A_\mathrm{BS}(u_1(t),u_2(t)) \approx (1-\omega) A_\mathrm{SS}(u_1) + \omega\,,
\end{equation}
so that one approaches the light curve for a blended single source which now is
object 1, and $\omega \approx 1-f$.
For $u = 2$, one obtains
$A_\mathrm{SS} = \frac{3}{2\sqrt{2}} \approx 1.06$, so that for $u_\mathrm{min,1}
\ge 2$ ($u_\mathrm{min,2} \ge 2$), the light curves for a blended single
source object 2 (1) and for a binary source differ by less than 6~\%. 
This shows that {\em any} blended event will have a successful fit 
with a binary source, where large
uncertainties in some of the fit parameters are expected, because the
binary source model involves more parameters than that for a blended single
source, namely for $n$ spectral bands the distance parameter $u_\mathrm{min,1}$ ($u_\mathrm{min,2}$),
the point of time $t_\mathrm{max,1}$ ($t_\mathrm{max,2}$) and the
luminosity offset ratios $\omega_i$ (altogether $n+2$ parameters) 
are convolved into $n$ blending parameters $f_i$.

Griest \& Hu (\cite{grihu}) have performed a comprehensive study on the 
types of events which arise for binary sources. Depending on the type of
the primary star of the binary system and the lens mass, they find that
among the binary source events, 60--95~\% have a light curve which is mainly
effected by one of the objects only. They call these events ``offset bright''
event if this object is the brighter one and ``offset dim'' or ``merged offset
dim'' of this object is the dimmer one\footnote{In contrast to
``merged offset dim'' events, there exist
two disjoint regions 
(near the two binary source objects) where the amplification is larger
than the detection threshold $A_\mathrm{T}$ (usually defined as $A_\mathrm{T} = 1.34$)
for ``offset dim'' events.}, where 50--80~\% of the binary source
events fall into the category ``offset bright'', 7--20~\% into the category
``offset dim'' and 0.3--2~\% into the category ``merged offset dim''.
The large fraction of these types of events among the binary source events
means that it is likely that events due to binary sources
can be successfully fitted with the model of a blended single source.

\section{OGLE\#5 as an example}
The points mentioned above can be illustrated using the event OGLE\#5 as an example.
Table~\ref{o5blend} shows the result of a fit for
a single source with and without blending,
while Table~\ref{o5binary} shows the result of a fit with a binary source.

\begin{table}[htb]
\caption[ ]{OGLE \#5: Fits for a single source with and without blending}
\label{o5blend}
\begin{flushleft}
\begin{tabular}{lcc}
\hline\noalign{\smallskip}
parameter & no blending & blending \\
\noalign{\smallskip}\hline\noalign{\smallskip}
\rule[-1ex]{0ex}{3.5ex}$t_\mathrm{E}$ [d]& 
$12.48_{-0.36}^{+0.36}$ & 
$62_{-10}^{+14}$ \\ 
\rule[-1ex]{0ex}{3.5ex}$t_\mathrm{max}$ [d]&
$824.331_{-0.017}^{+0.017}$ &
$824.36_{-0.017}^{+0.018}$ \\ 
\rule[-1ex]{0ex}{3.5ex}$u_\mathrm{min}$ & 
$0.0848_{-0.0013}^{+0.0013}$ &
$0.0137_{-0.0025}^{+0.0027}$ \\
\rule[-1ex]{0ex}{3.5ex}$m_\mathrm{base}$ &
$-17.904_{-0.011}^{+0.011}$ &
$-17.960_{-0.013}^{+0.012}$ \\ 
\rule[-1ex]{0ex}{3.5ex}$f$ &
--- &
$0.166_{-0.030}^{+0.032}$ \\ \noalign{\smallskip}
\hline\noalign{\smallskip}
\rule[-1ex]{0ex}{3.5ex}$\chi^2_\mathrm{min}$ & 
361.00 &
117.93 \\ 
\rule[-1ex]{0ex}{3.5ex} \# d.o.f = $n$ & 
97 &
96 \\ 
\rule[-1ex]{0ex}{3.5ex}$\sqrt{2\chi^2_\mathrm{min}}-\sqrt{2n-1}$ &  
12.99 &
1.537 \\ 
\rule[-1ex]{0ex}{3.5ex}$P(\chi^2 \geq \chi^2_\mathrm{min})$ & 
$8\cdot 10^{-39}$ &
6~\% \\
\noalign{\smallskip} \hline
\end{tabular}
\end{flushleft}
\end{table} 

\begin{table}[htb]
\caption[ ]{OGLE \#5: Binary source fit}
\label{o5binary}
\begin{flushleft}
\begin{tabular}{lc}
\hline\noalign{\smallskip}
parameter & OGLE\#5 \\\noalign{\smallskip}
\hline 
\rule[-1ex]{0ex}{3.5ex}$t_\mathrm{E}$ [d]& 
$26.3_{-3.6}^{+19}$  \\ 
\rule[-1ex]{0ex}{3.5ex}$t_\mathrm{max,1}$ [d]&
$824.359_{-0.018}^{+0.018}$ \\
\rule[-1ex]{0ex}{3.5ex}$t_\mathrm{max,2}$ [d]&
$827.1_{-3.4}^{+73}$ \\ 
\rule[-1ex]{0ex}{3.5ex}$u_\mathrm{min,1}$ & 
$0.032_{-0.013}^{+0.005}$ \\ 
\rule[-1ex]{0ex}{3.5ex}$u_\mathrm{min,2}$ & 
$0.89_{-0.19}^{+0.28}$ \\ 
\rule[-1ex]{0ex}{3.5ex}$\omega$ & 
$0.618_{-0.054}^{+0.15}$ \\ 
\rule[-1ex]{0ex}{3.5ex}$m_\mathrm{base}$ &
$-17.958_{-0.012}^{+0.012}$ \\ 
\noalign{\smallskip}\hline\noalign{\smallskip}
\rule[-1ex]{0ex}{3.5ex}$\chi^2_\mathrm{min}$ & 
110.58 \\ 
\rule[-1ex]{0ex}{3.5ex} \# d.o.f = $n$ & 
94 \\ 
\rule[-1ex]{0ex}{3.5ex}$\sqrt{2\chi^2_\mathrm{min}}-\sqrt{2n-1}$ & 
1.197  \\ 
\rule[-1ex]{0ex}{3.5ex}$P(\chi^2 \geq \chi^2_\mathrm{min})$ & 
12~\% \\
 \noalign{\smallskip}\hline
\end{tabular}
\end{flushleft}
\end{table}

For the fits, amplification values have been used rather than the magnification
values as obtained from the OGLE collaboration. These amplification values
refer to a baseline, which has been obtained
by fitting the tail region to a 
constant brightness. One obtains
also a scaling factor $\gamma$, which corresponds to the most-likely
size of the errors. Table~\ref{o16resc} shows the results.

\begin{table}[htb]
\caption[ ]{OGLE \#5: fixing of the baseline and scaling factor $\gamma$}
\label{o16resc}
\begin{flushleft}
\begin{tabular}{lc}
\hline\noalign{\smallskip}
parameter & value \\
\noalign{\smallskip}\hline\noalign{\smallskip}
\rule[-1ex]{0ex}{3.5ex}begin peak [d]&
775 \\ 
\rule[-1ex]{0ex}{3.5ex}end peak [d]&
875 \\ 
\rule[-1ex]{0ex}{3.5ex}$m_\mathrm{base}$ & 
-17.9524 \\ 
\rule[-1ex]{0ex}{3.5ex}$\gamma$ & 
1.503 \\ \noalign{\smallskip}\hline 
\end{tabular}
\end{flushleft}
\end{table} 

For the fits of the
light curve, the errors have been increased by the factor $\gamma$.
The need for rescaling arises from the fact that the assumption of a
constant tail does not hold for the original data. 
The error bounds shown correspond to projections of the hypersurface
$\Delta \chi^2 = \chi^2-\chi^2_\mathrm{min} = 1$.

Note that the single source fit without blending is not acceptable.
The error bounds on $t_\mathrm{E}$ are large for the fit with blending: 
The boundaries of the
$1$-$\sigma$-intervals
differ by a factor of 1.5 
so that the expectation values for the masses would differ by a factor
of about~2.

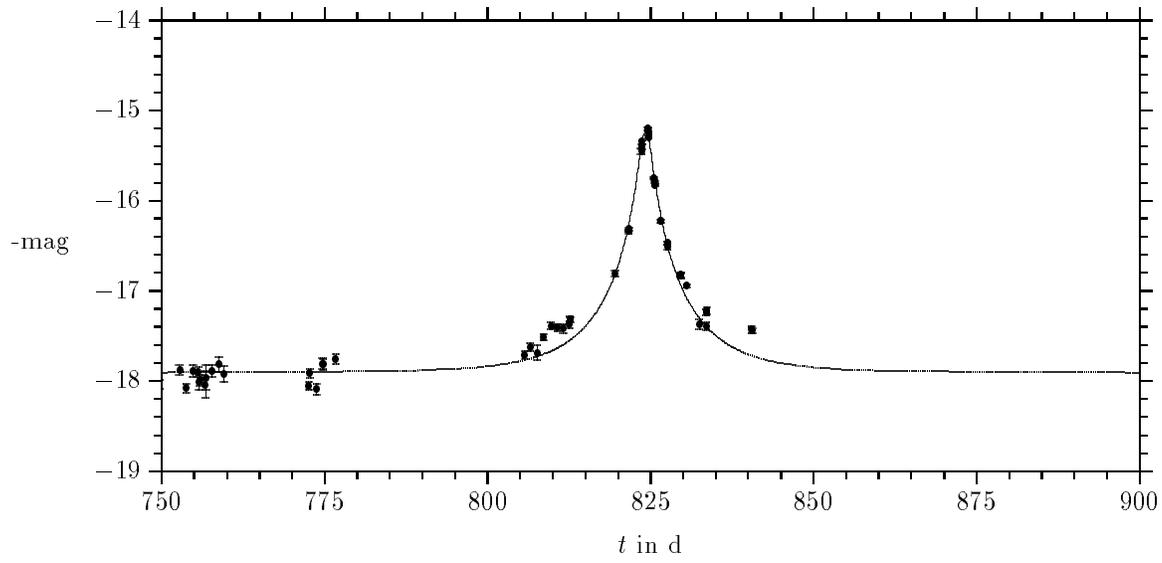
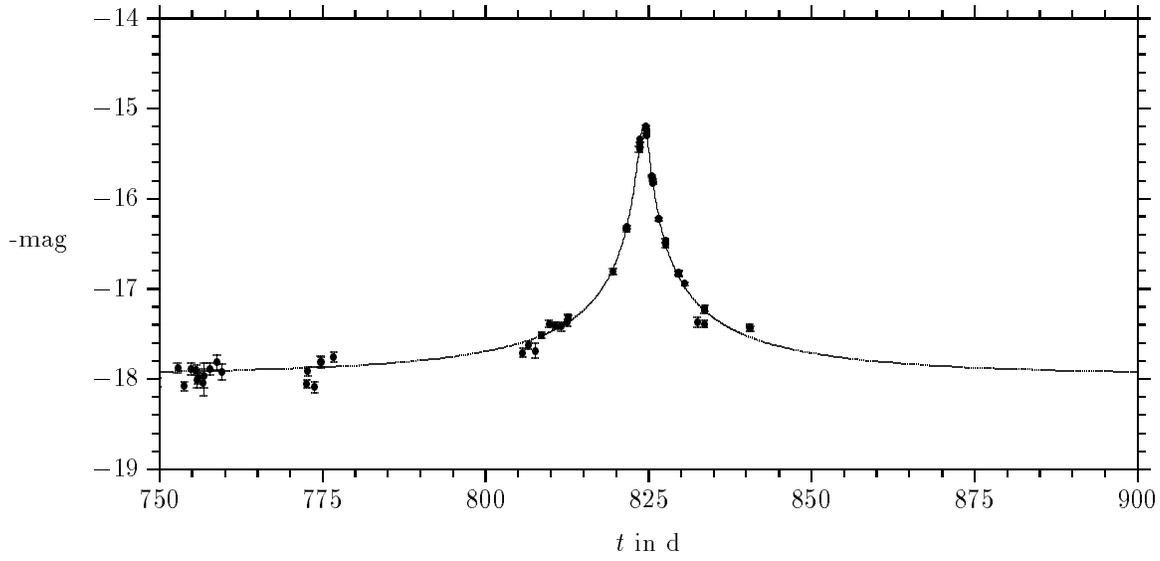
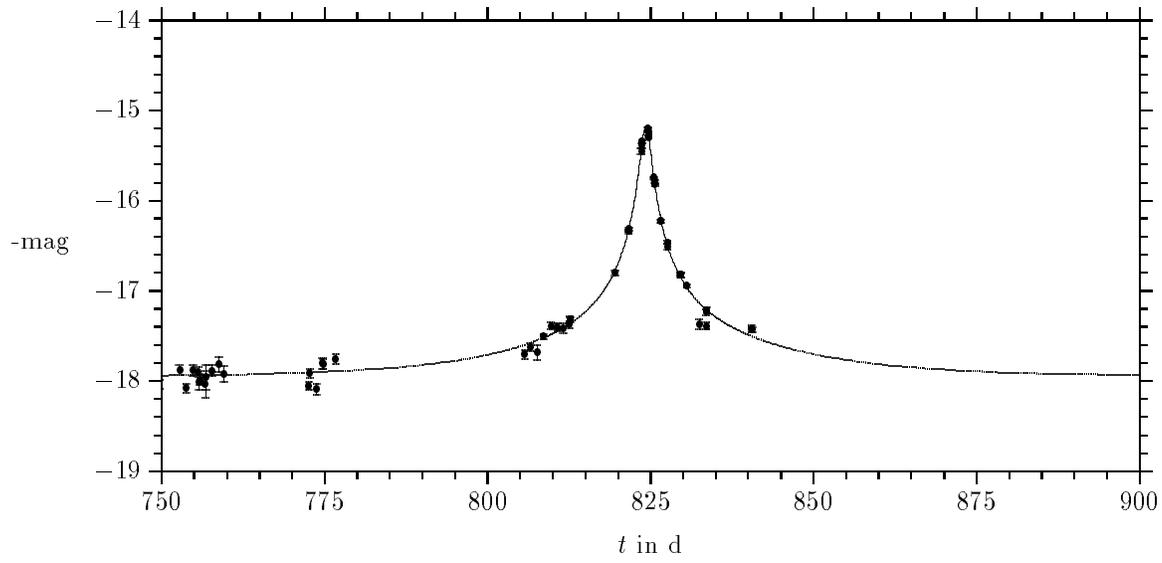
\begin{figure*}
\vspace{21cm}
\caption{OGLE\#5: Fit with a single source (top), a blended single source 
(middle) and a binary source (bottom).}
\label{o5lc}
\end{figure*}

The differences between the fits with and without blending can be seen
in the light curves of the peak region in Fig.~\ref{o5lc}.
The magnitudes are shown as the ordinate, which
allows to see the data and the light curve in the peak better, though the
fits have been performed using the amplification values. 
One sees a dramatic improvement of the fit with blending
compared with fit without. For the fit without blending, 
one has many discrepant points in the
wings of the light curve, which is not the case for the fit with blending.

The light curve for a binary source (Fig.~\ref{o5lc}) is similar to
that for a single source with blending.
The lens passes close to object 1, whereas the minimal separation to the
position of object 2 projected onto the lens plane is about $0.7$--$1.2$~$r_\mathrm{E}$.
The luminosity offset ratio is $\omega$ in agreement with the blending
parameter $1-f$ if one considers the quoted $1$-$\sigma$-bounds --- so that the binary source corresponds to a single source object 1 event
with blending ---, the values
meet at about $1.2 \sigma$. For the 
binary source fit, there are large 
uncertainties in the event time scale $t_\mathrm{E}$, the minimal distance
to object 2, $u_\mathrm{min,2}$, the time of closest approach to object 2,
$t_\mathrm{max,2}$, and in the luminosity offset ratio $\omega$. 
The extremely large upper bound on $t_\mathrm{max,2}$ is due to the lack of
data points for $t > 840~\mbox{d}$, since light curves are possible which
involve another peak in this region. However, this bound is not arbitrarily
large because this peak still has some influence to the right wing of the peak near 
$t = 825~\mbox{d}$. If this peak would move to infinity one would approach the
fit for a single source with blending which however 
has a $\chi^2_\mathrm{min}$ which is larger by about 7.

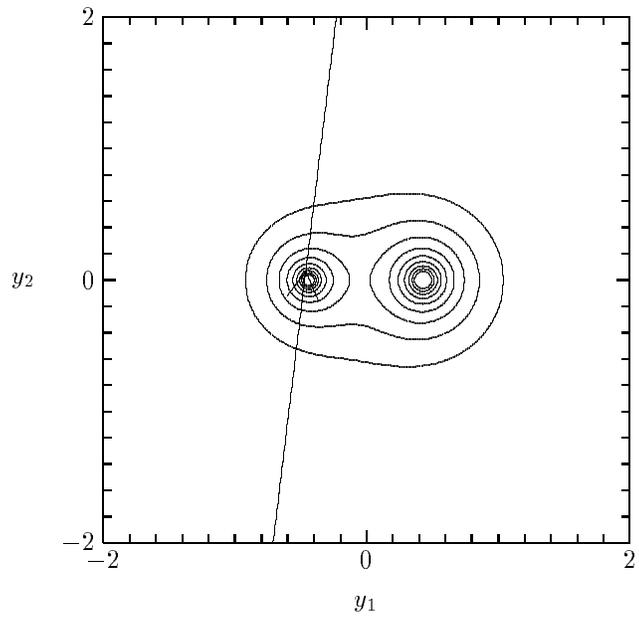
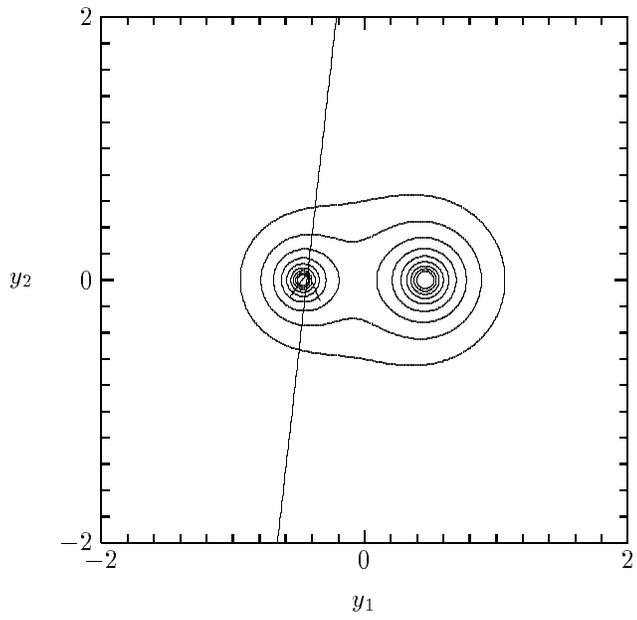
\begin{figure*}
\vspace{8.5cm}
\caption{OGLE\#5: Magnification contour plot for the binary source fit and
the cis-configuration (left) and the trans-configuration (right) 
together with the lens trajectory. 9 contours with
$\Delta \mathrm{mag} = 0.5\ldots{}2.5$ in steps of 0.25.} 
\label{o5cfbs}
\end{figure*}

Figure~\ref{o5cfbs} shows the lens 
trajectory and the magnification
contours for the binary source fit for both the cis- and the trans-configuration.
All distances are measured in Einstein radii projected to the source plane
$r_\mathrm{E}'$.

This example demonstrates that a blended event can successfully be
explained by a binary source model and that the error bounds on the parameters
which are convolved into the blending parameter are large. However, since
the binary source model corresponds to a ``merged offset dim'' event which
should not occur frequently (around 1~\% of the binary source events) it does
not show the other direction clearly.

\section{Summary and conclusions}
I have shown that any event involving a blended single source can also be
explained by a (non-blended) binary source and argued 
that most of the binary source events could be successfully explained 
by a blended single source. This may explain the lack of events which 
have been claimed as being due to binary sources.
This also shows, that due to the intrinsic similarity of light curves of
blended single sources and non-blended binary sources, it is difficult to
distinguish between these effects.

\begin{acknowledgements}
I would like to thank A.~C.~Hirshfeld, S.~Mao, and C.~Alard  for reading 
different versions of the manuscript, the OGLE collaboration for making available their data
and an anonymous referee for commenting on some points which helped to clarify 
them.
\end{acknowledgements}

\clearpage

\begin{figure*}
{\LARGE Figure 1 (top)}\\[5mm]

\epsfig{file=h0524.f1a}
\end{figure*}

\clearpage

\begin{figure*}
{\LARGE Figure 1 (middle)}\\[5mm]

\epsfig{file=h0524.f1b}
\end{figure*}

\clearpage
\begin{figure*}
{\LARGE Figure 1 (bottom)}\\[5mm]

\epsfig{file=h0524.f1c}
\end{figure*}

\clearpage
\begin{figure*}
{\LARGE Figure 2 (left)}\\[5mm]

\epsfig{file=h0524.f2a}
\end{figure*}

\clearpage
\begin{figure*}
{\LARGE Figure 2 (right)}\\[5mm]

\epsfig{file=h0524.f2b}
\end{figure*}

\end{document}